\documentclass{article}

\usepackage[final,nonatbib]{nips_2018}

\usepackage{graphicx}
\usepackage{subfigure}
\graphicspath{ {./images/} }
\usepackage[utf8]{inputenc} 
\usepackage[T1]{fontenc}    
\usepackage{hyperref}       
\usepackage{url}            
\usepackage{booktabs}       
\usepackage{amsfonts}       
\usepackage{nicefrac}       
\usepackage{microtype}      
\usepackage{amsmath}
\usepackage[sorting=none]{biblatex}
\title{Forecasting NYC Yellow Taxi Ridership Decline: A Time Series Analysis of Daily Passenger Counts (2017-2019)}
\author{
   Gaurav Singh \\
   \texttt{grvsingh@g.ucla.edu}
}

\begin{document}

\maketitle

\vspace{-0.25cm}
\begin{abstract}
This study analyzes and forecasts daily passenger counts for New York City's iconic yellow taxis during 2017-2019, a period of significant decline in ridership. Using a comprehensive dataset from the NYC Taxi and Limousine Commission, we employ various time series modeling approaches, including ARIMA models, to predict daily passenger volumes. Our analysis reveals strong seasonal patterns, with a consistent linear decline of approximately 200 passengers per day throughout the study period. After comparing multiple modeling approaches, we find that a simple AR(1) model, combined with careful detrending and cycle removal, provides the most accurate predictions, achieving a test RMSE of 34,880 passengers on a mean ridership of 438,000 daily passengers. The research provides valuable insights for policymakers and stakeholders in understanding and potentially addressing the declining trajectory of NYC's yellow taxi service.
\end{abstract}

\texttt{The implementation for this research can be found here:\cite{repolink}}
\vspace{-0.35cm}
\section{Introduction}
New York City's yellow taxis have long been a defining symbol of urban transportation, yet their role in the city's mobility landscape is undergoing significant transformation. With approximately 13,500 medallion taxis operating a decade ago reduced to fewer than 9,000 by 2019, this decline reflects broader changes in urban transportation dynamics and consumer preferences.

As one of the world's premier metropolitan areas, New York City's transportation ecosystem serves millions of daily trips across its five boroughs. The yellow medallion taxi system, regulated by the New York City Taxi and Limousine Commission (TLC), has historically been the dominant form of for-hire transportation in the city, particularly in Manhattan \cite{taxitypes}. These vehicles are uniquely authorized to pick up street-hail passengers throughout Manhattan and at the city's airports, a privilege that once made medallion ownership extremely valuable.

The declining prevalence of yellow taxis can be attributed to several interconnected factors \cite{nypost}. The emergence of ride-hailing platforms such as Uber and Lyft has introduced dynamic pricing and convenient mobile booking options that appeal to modern consumers. Additionally, the introduction of green borough taxis in 2013 has created new competition in outer borough areas. Yellow taxi operators face particular challenges due to their fixed fare structures, which may not adequately respond to market conditions, and higher operational costs compared to ride-hailing drivers.

This study aims to analyze daily passenger count patterns for NYC yellow taxis from 2017 to 2019, identifying and quantifying temporal patterns, including seasonal cycles and long-term trends. Through the development and evaluation of time series models for predicting daily passenger counts, we seek to assess the sustainability of the current yellow taxi system based on ridership trends. The research focuses specifically on the pre-pandemic period to establish baseline patterns unaffected by the extraordinary circumstances of COVID-19.

Understanding these patterns and developing accurate forecasting models serves multiple stakeholders in the urban transportation ecosystem. Policy makers can utilize these insights to evaluate regulatory frameworks and intervention needs, while taxi fleet operators can better optimize their resource allocation. The findings also provide valuable input for urban planners assessing transportation system capacity requirements and economic analysts evaluating the taxi industry's financial sustainability. This research contributes to the broader discourse on urban transportation evolution and the future of traditional taxi services in modern cities.

\section{Dataset}
The dataset used is provided by NYC Taxi and Limousine Commission \cite{data} and is open-sourced. The data contains number of rides completed by a Taxi and the number of passengers corresponding to the trip is reported by the taxi driver. The dataset is provided in chunks on monthly basis for every trip. This project uses data from 01/01/2017 to 12/31/2019 i.e. 36 months of data. The whole data is merged into a single file and grouped by date to aggregate the daily passenger counts. Full consolidated dataset has 1095 days of data out of which last 61 days (11/01/2019 to 12/31/2019) are kept for testing and is not used for any analysis. COVID pandemic time (2020 on-wards) is not taken into consideration given at that time there were only 700 taxis in service.

\section{Exploratory Data Analysis}
\begin{figure}[h!]
  \centering
  \includegraphics[scale=0.35]{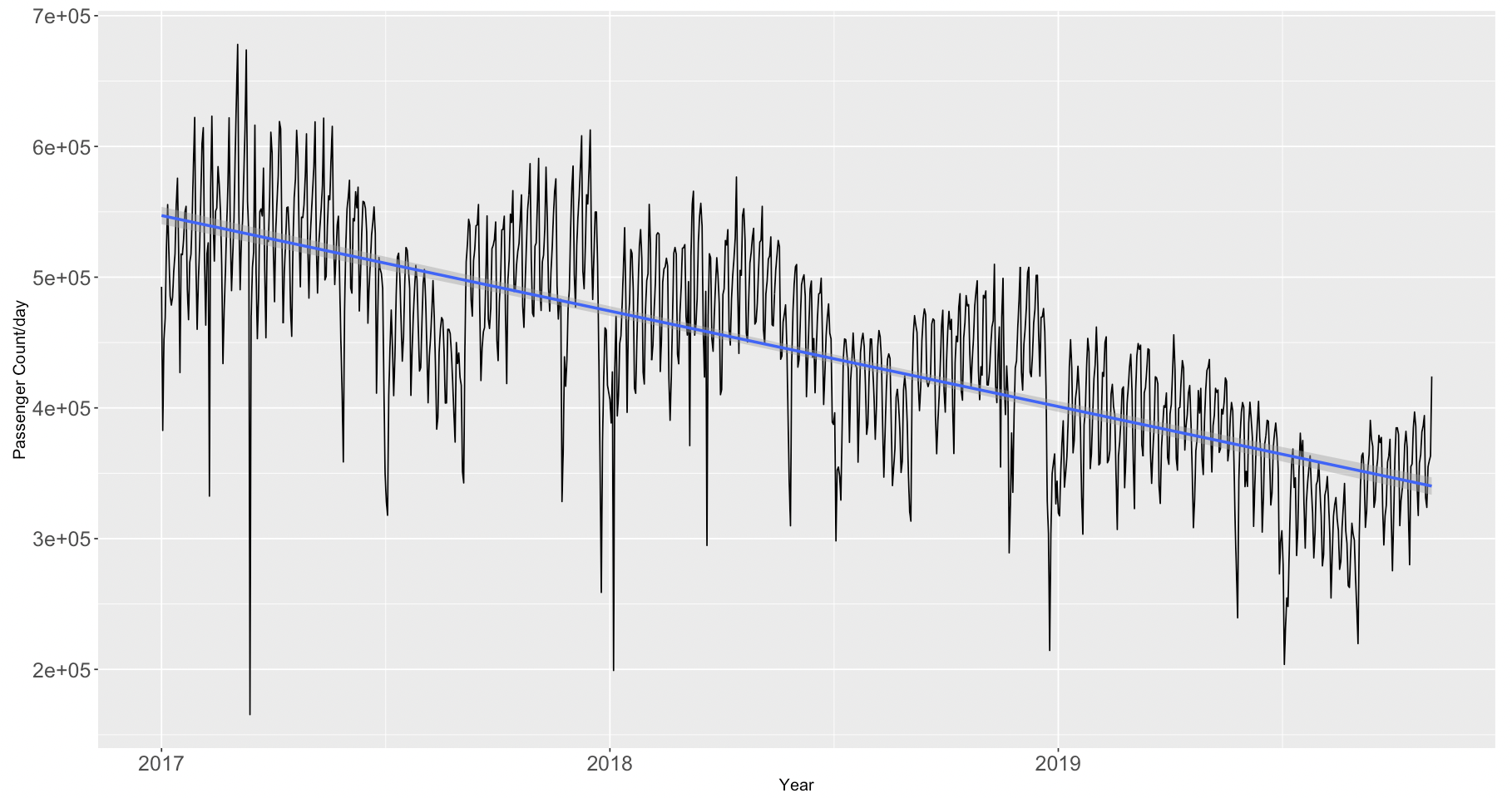}
  \caption{Raw Train Data + Linear trend}
  \label{fig:orig}
\end{figure}

Looking at the time series data from Fig.~\ref{fig:orig}, it looks like data is quite noisy, discontinuous with a strong presence of cycles and seasonality. This is evident because repeating patterns are observed in the plot. It's observed that overall trend of the data is decreasing and a linear trend would be a good estimate for this decrease which is shown using the blue line in Fig.~\ref{fig:orig}. The fitted trend has a slope of -200.13 and a intercept of 547,356 as shown in Table ~\ref{tab:lm}, which shows that in the beginning of 2017 there were $\sim 550K$ daily passengers but the mean of the train data is $\sim 438K$ which shows that over the time the passengers are decreasing, precisely around 200 passengers/day.

\begin{figure}[h!]
    \centering
    \subfigure[]{\includegraphics[scale=0.25]{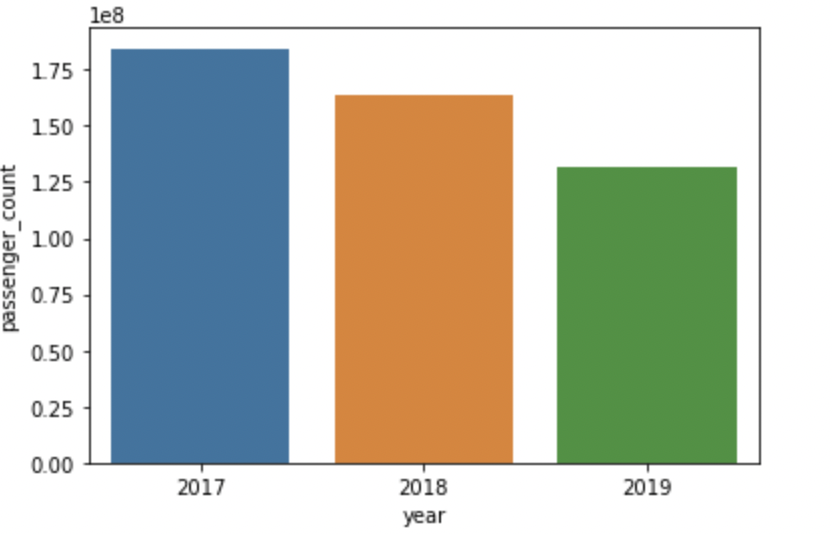}} 
    \hspace{0.5cm}
    \subfigure[]{\includegraphics[scale=0.25]{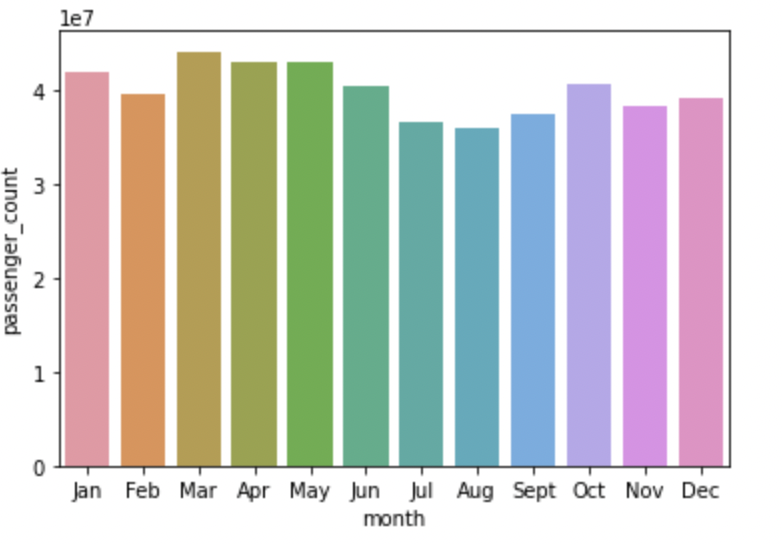}}
    \hspace{0.5cm}
    \subfigure[]{\includegraphics[scale=0.25]{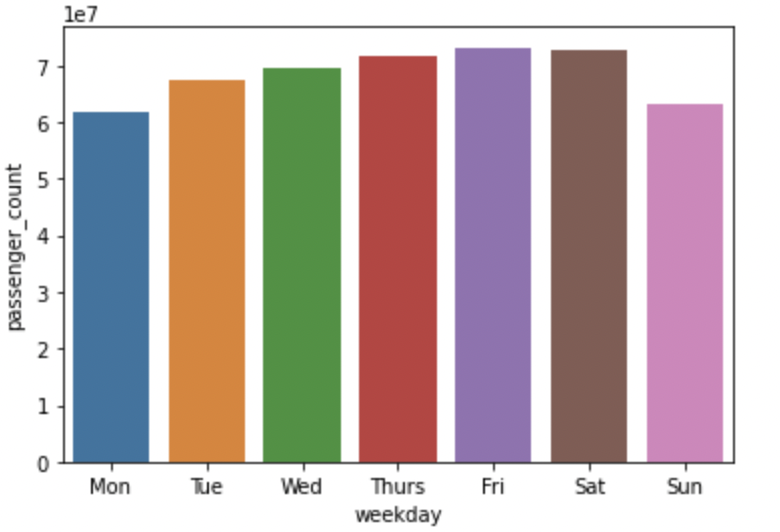}}
    \caption{Aggregated Passenger Counts (a) Yearly (b) Monthly (c) Weekday}
    \label{fig:agg}
\end{figure}

\begin{figure}[h!]
    \centering
    \subfigure[]{\includegraphics[scale=0.40]{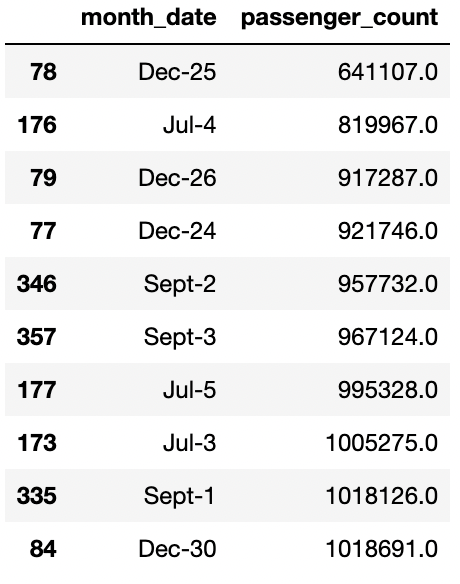}} 
    \hspace{0.5cm}
    \subfigure[]{\includegraphics[scale=0.40]{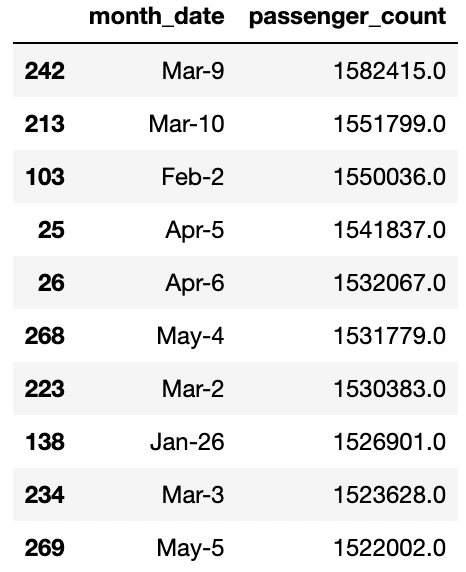}}
    \caption{Top-10 Dates with (a) Least passengers  (b) Most passengers}
    \label{tab:dates}
\end{figure}

Fig.~\ref{fig:agg}(a) shows the total yearly passenger counts and the decay is quite linear from 2017 to 2019 which supports the choice of linear trend. Monthly aggregates from Fig.~\ref{fig:agg}(b), shows that Spring season is favourable for Yellow taxis with large number of passengers as compared to Winter season where the counts are quite scarce. This makes sense because NYC sees quite cold winters with snowfall which restricts people from going out. Weekly aggregates from Fig.~\ref{fig:agg}(c) indicates that Monday and Sunday are not favourable for taxi business whereas there is a linear growth in passengers as the week progresses from Tuesday to Saturday. Further inspection on the individual dates shows that the mid-March week which marks the beginning of Spring in NYC seems to be most profitable for taxi drivers and Holiday weeks like Christmas and Independence week seems to be least profitable with lowest no. of passengers as shown in Table ~\ref{tab:dates}

\begin{figure}[h!]
  \centering
  \includegraphics[scale=0.4]{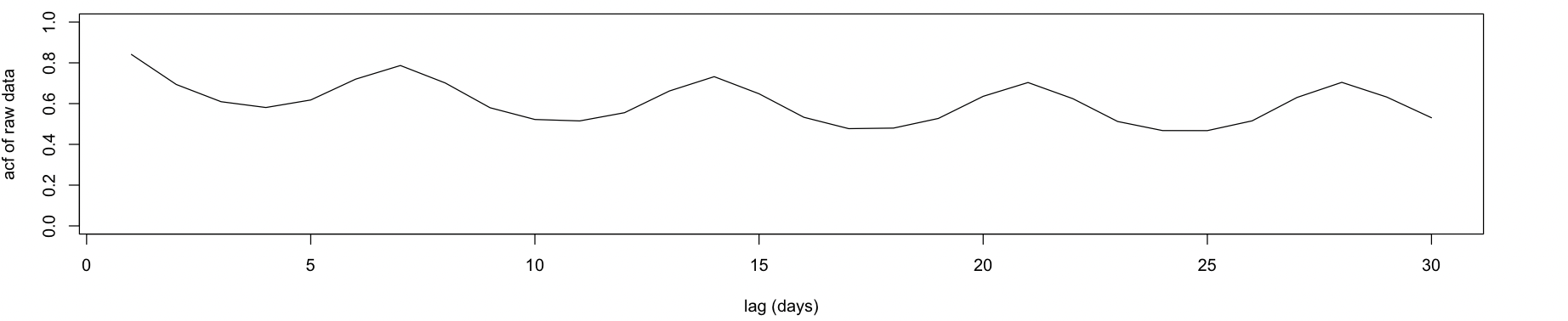}
  \caption{ACF plot for raw data}
  \label{fig:acf_o}
\end{figure}

\begin{figure}[h!]
  \centering
  \includegraphics[scale=0.4]{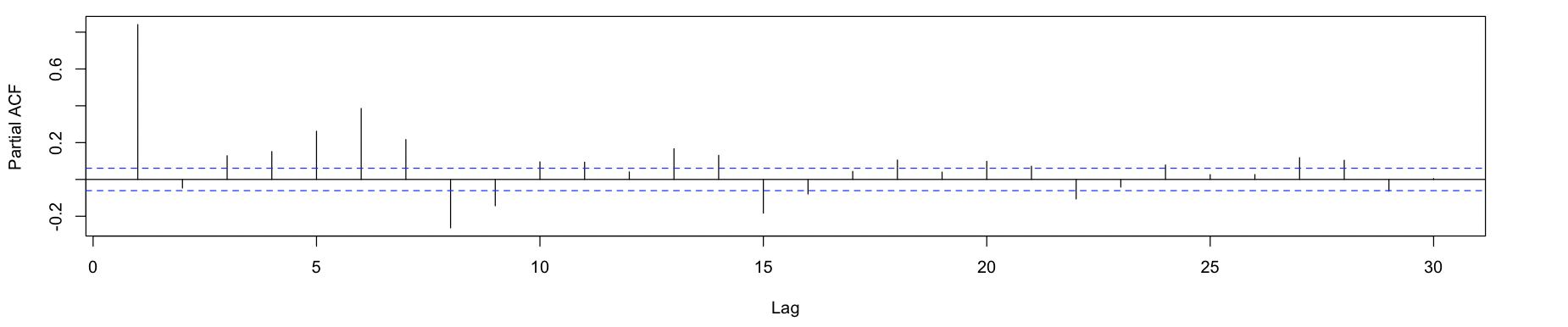}
  \caption{PACF plot for raw data}
  \label{fig:pacf_o}
\end{figure}

From the ACF plot in Fig.~\ref{fig:acf_o}, it looks like ACF is oscillating and gradually decreasing whereas the PACF in Fig.~\ref{fig:pacf_o} seems to cut off at a lag 2, which hints that this time series can be modelled well by an AR(1) process. Each lag in the ACF and PACF plots represents 1 day in a year.

\section{Data Preparation before Time Series Modelling}
\begin{figure}[h!]
  \centering
  \includegraphics[scale=0.4]{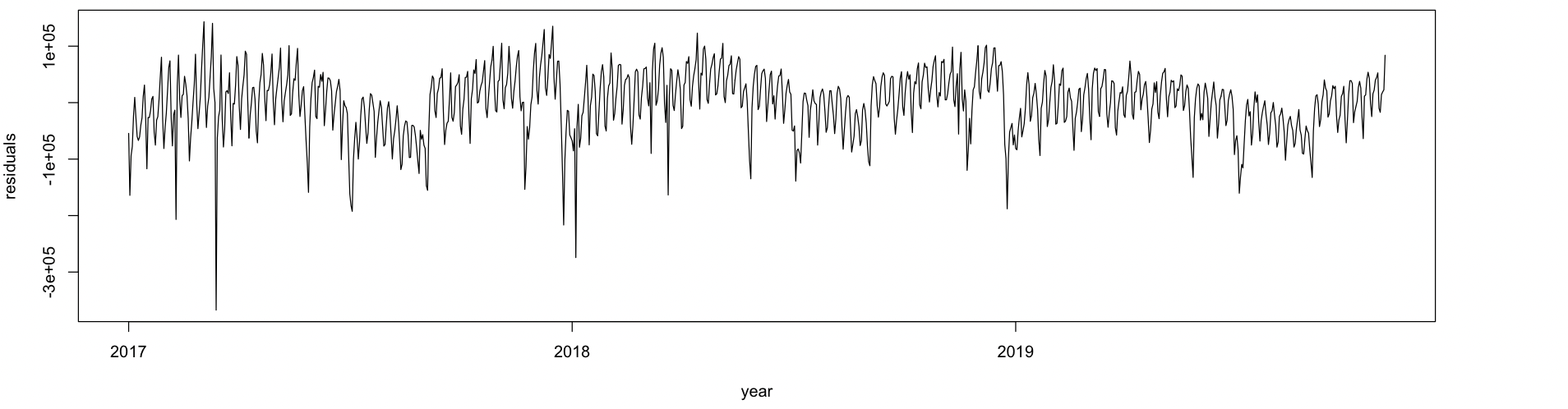}
  \caption{Residuals after linear detrending}
  \label{fig:detrended}
\end{figure}

\begin{figure}[h!]
  \centering
  \includegraphics[scale=0.4]{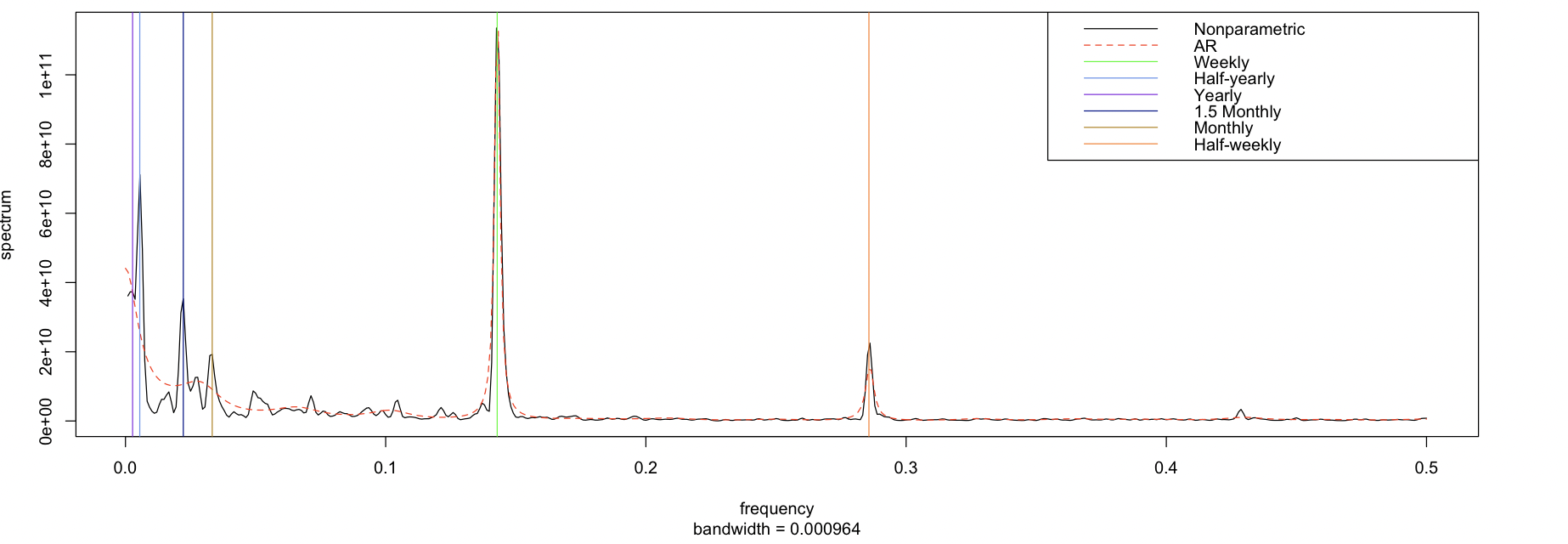}
  \caption{Frequency densities for detrended residuals}
  \label{fig:spectrum}
\end{figure}

Before any time series modelling it is absolutely necessary that it's almost stationary and that it's free from any dominating trends and cycles. Augmented Dickey-Fuller Test conducted on train data has a p-value of 0.01 which shows that despite the series being discontinuous with strong presence of cycles and trends, it's Stationary with low confidence. The first thing is to remove the linear trend from the train data which gave the residuals as shown in Fig.~\ref{fig:detrended}. Looking at the residuals, it looks much more stationary but the cyclic patterns are still intact in it. To identify such cycles, AR spectral estimation and Daniell window (span = 3), smoothened Spectral analysis is conducted and the Frequency densities are plotted as shown in Fig.~\ref{fig:spectrum}.

\begin{figure}[h!]
    \centering
    \subfigure[]{\includegraphics[scale=0.15]{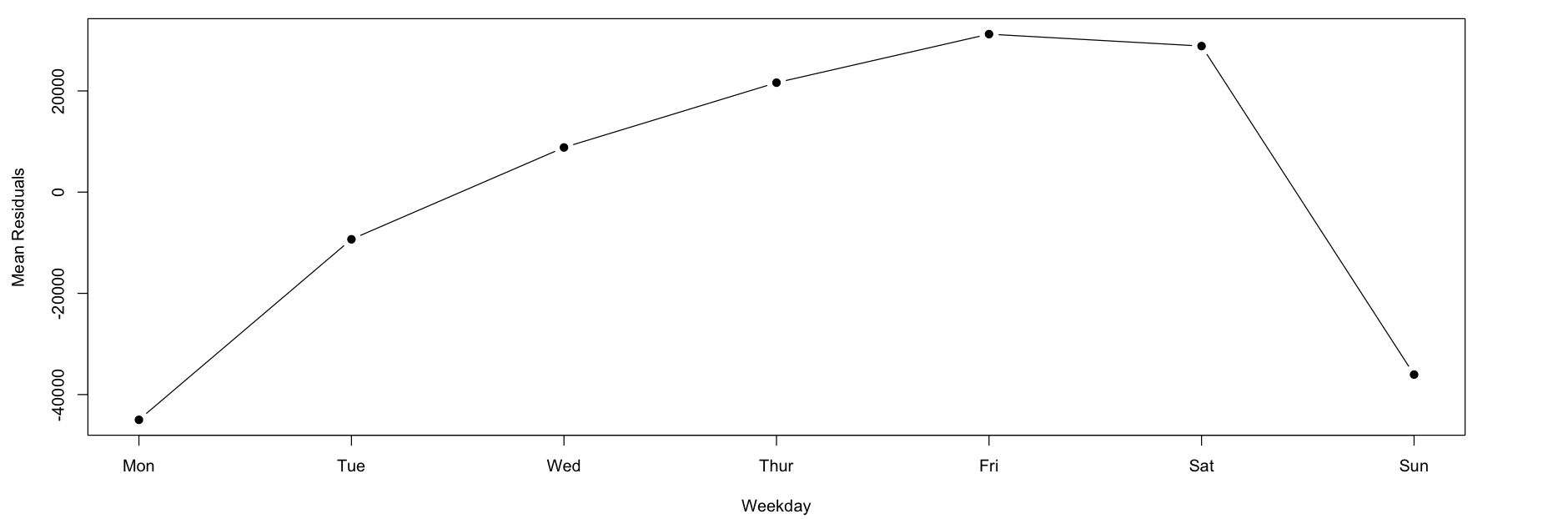}} 
    \hspace{0.5cm}
    \subfigure[]{\includegraphics[scale=0.15]{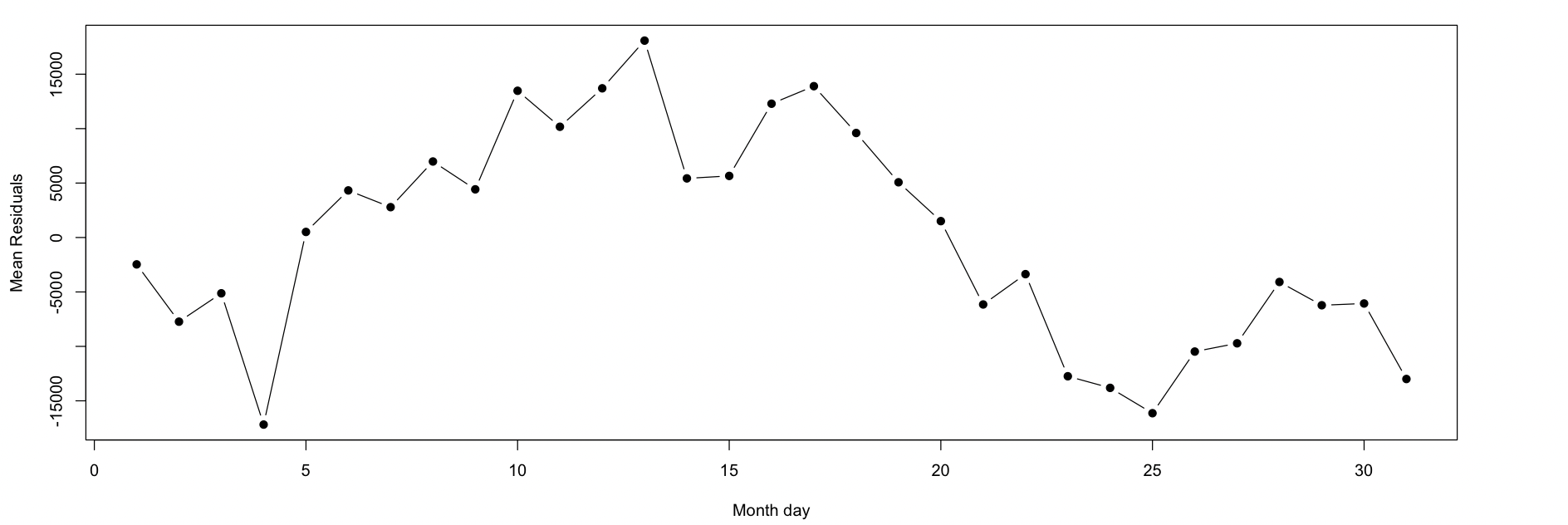}}
    \hspace{0.5cm}
    \subfigure[]{\includegraphics[scale=0.15]{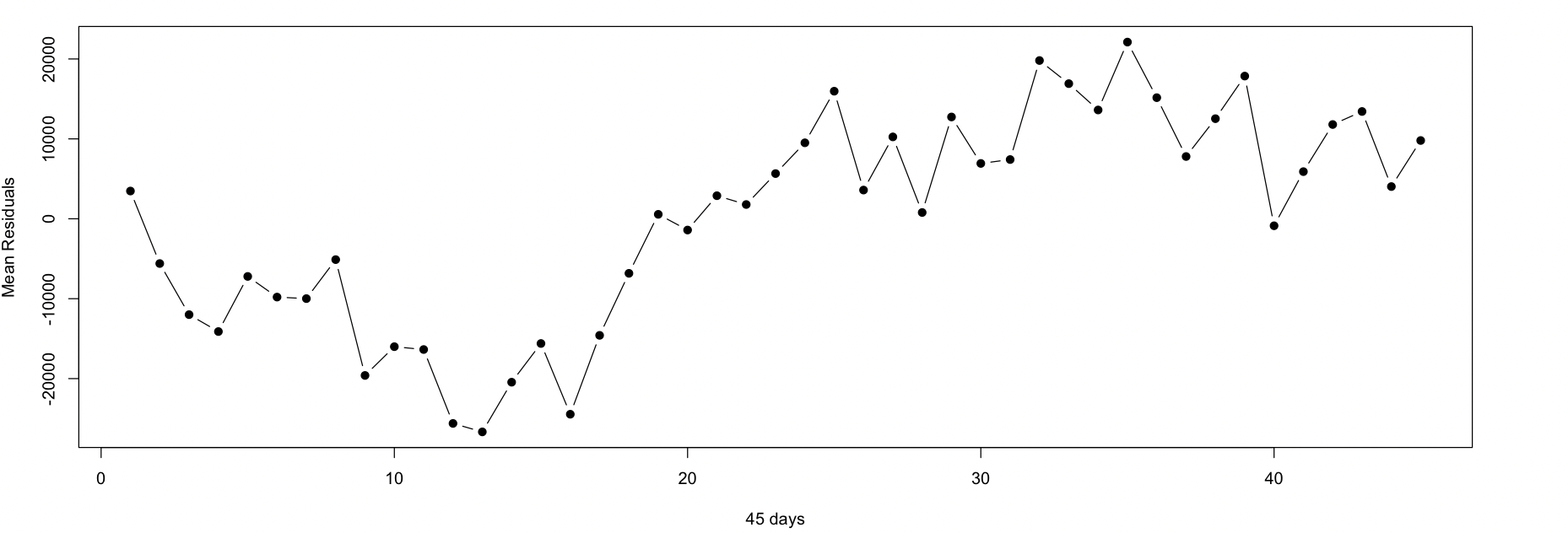}}
    \hspace{0.5cm}
    \subfigure[]{\includegraphics[scale=0.15]{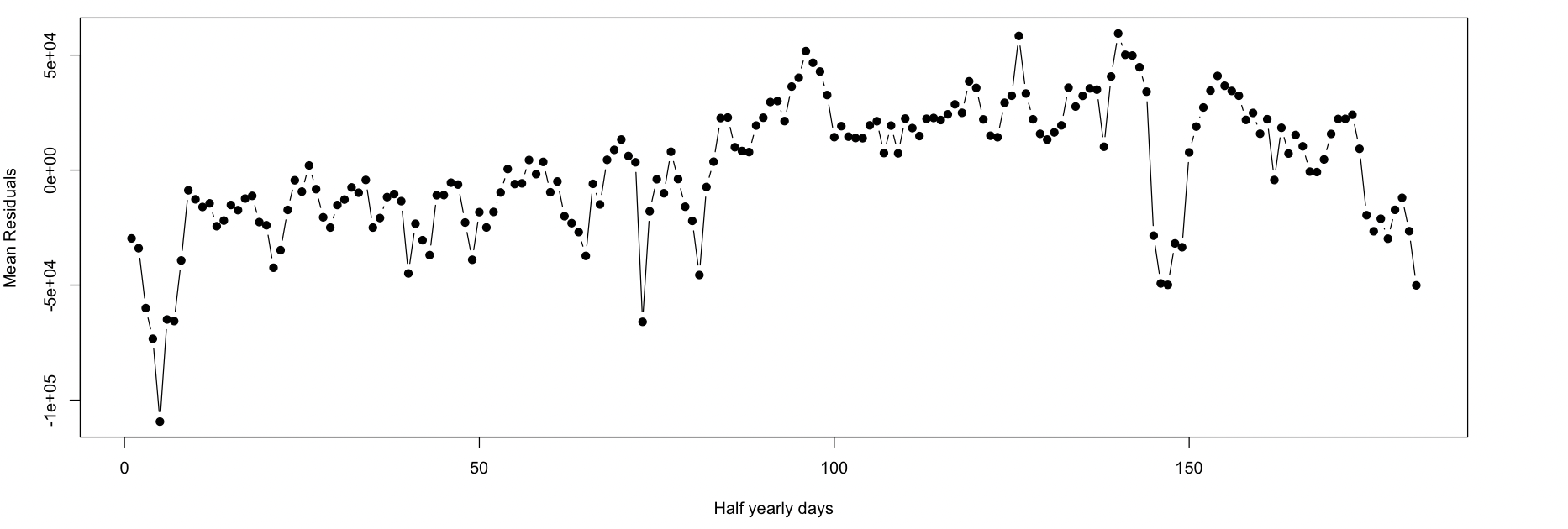}}
    \hspace{0.5cm}
    \subfigure[]{\includegraphics[scale=0.15]{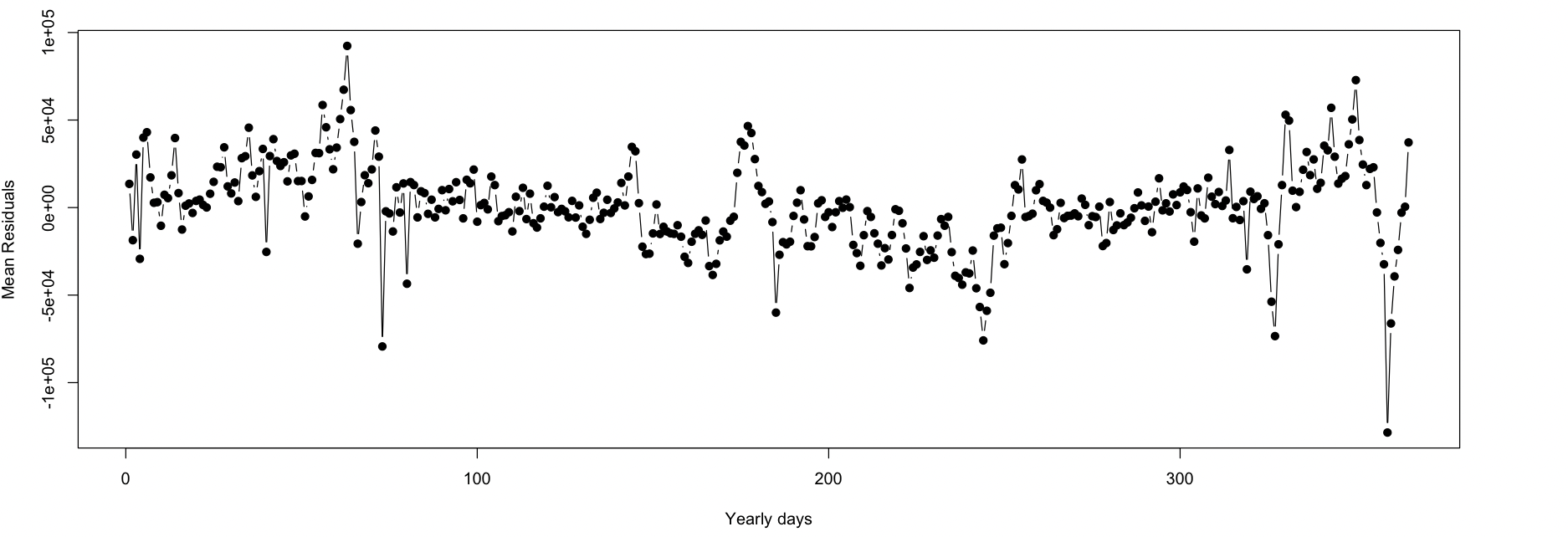}}
    \hspace{0.5cm}
    \caption{(a) Weekly cycle  (b) Monthly cycle (c) 45-day cycle (d) Half-yearly cycle (e) Yearly cycle}
    \label{fig:cycleplot}
\end{figure}

From the both the AR and non-parametric spectrum, it's evident that there is a presence of lot of low frequency (large period) cycles, but the major concern is with the sharp sudden spikes in the plot which carry too much power and dominate the time series. 6 such cycles are identified with weekly cycle(7 day) carrying the most power, followed by Half-yearly(182 day), Yearly(365 day), One and a half month (45 day), Monthly(30 day) and Half-weekly cycles (in decreasing order of power contribution). Out of these 6 dominant cycles, 5 cycles having most power are removed from the residuals as shown in Fig.~\ref{fig:cycleplot}. The method opted for multiple cycle removal is the removal of mean of the corresponding days of cycle from the residuals. The cycles are removed sequentially one at a time and the means are computed only from the train data residuals left after each successive removal. It is observed that removing a large period cycle before a shorter period cycle introduced some new cycles whose period is hard to estimate. This was not the case if shorter cycles are removed before the longer ones. The 5 cycles are shown in Fig. \ref{fig:cycleplot} where mean residuals are plotted for each day. The weekly cycle resembles in shape with the total weekly distribution shown in Fig. \ref{fig:agg}(c).

\begin{figure}[h!]
  \centering
  \includegraphics[scale=0.4]{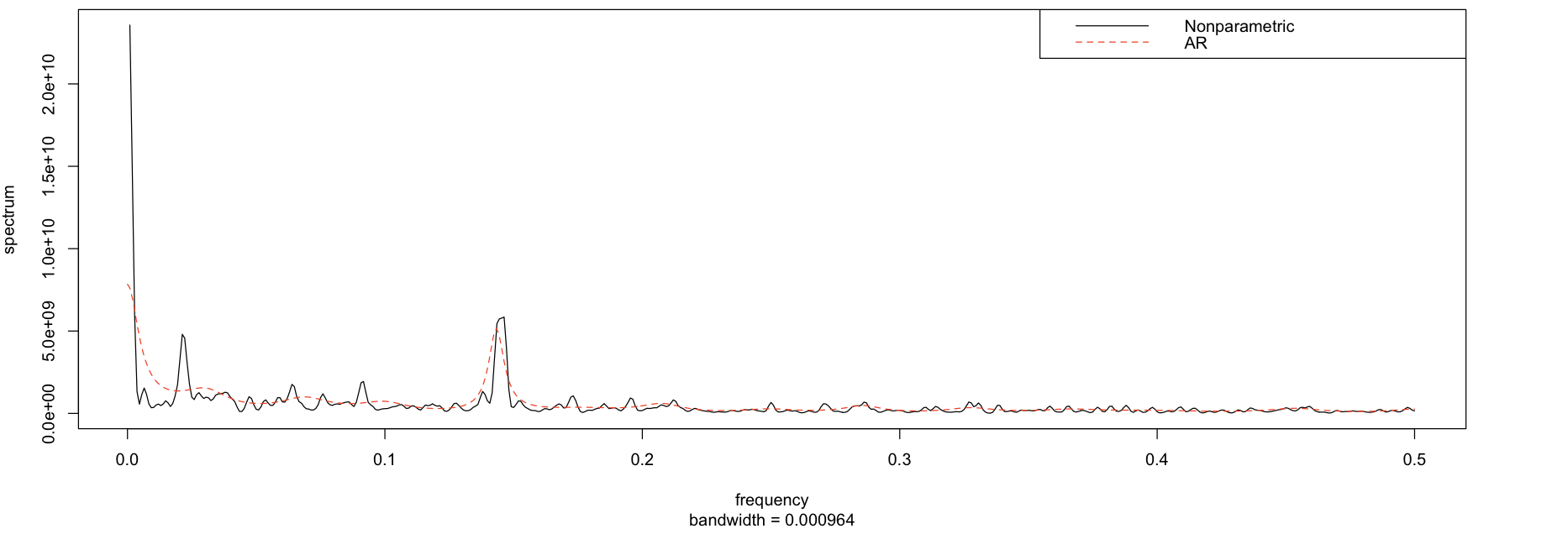}
  \caption{Frequency densities for cycle removed residuals}
  \label{fig:cycremov}
\end{figure}

\begin{figure}[h!]
  \centering
  \includegraphics[scale=0.4]{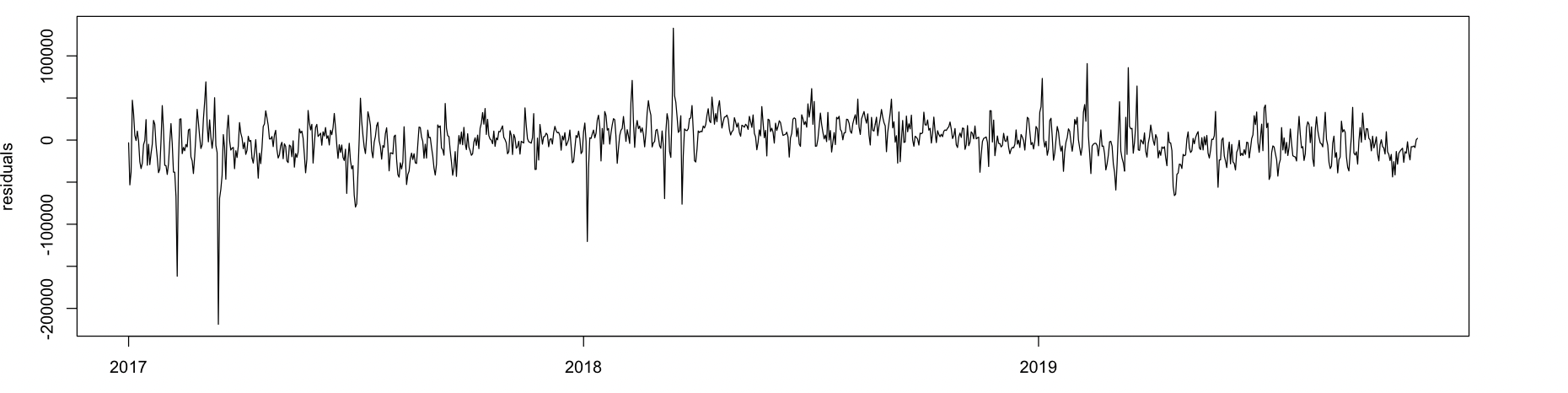}
  \caption{Final Residuals}
  \label{fig:fresid}
\end{figure}

\begin{figure}[h!]
  \centering
  \includegraphics[scale=0.4]{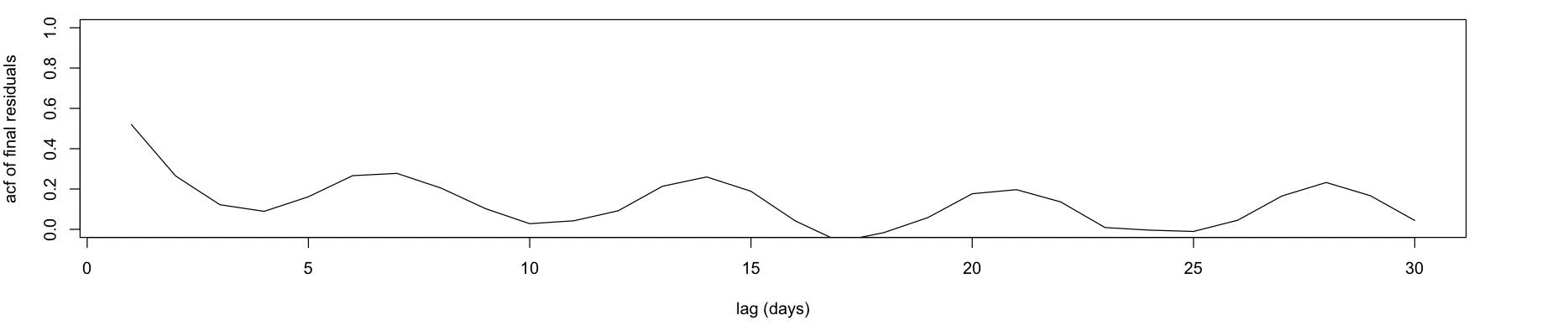}
  \caption{Final Residuals: ACF}
  \label{fig:facf}
\end{figure}

\begin{figure}[h!]
  \centering
  \includegraphics[scale=0.4]{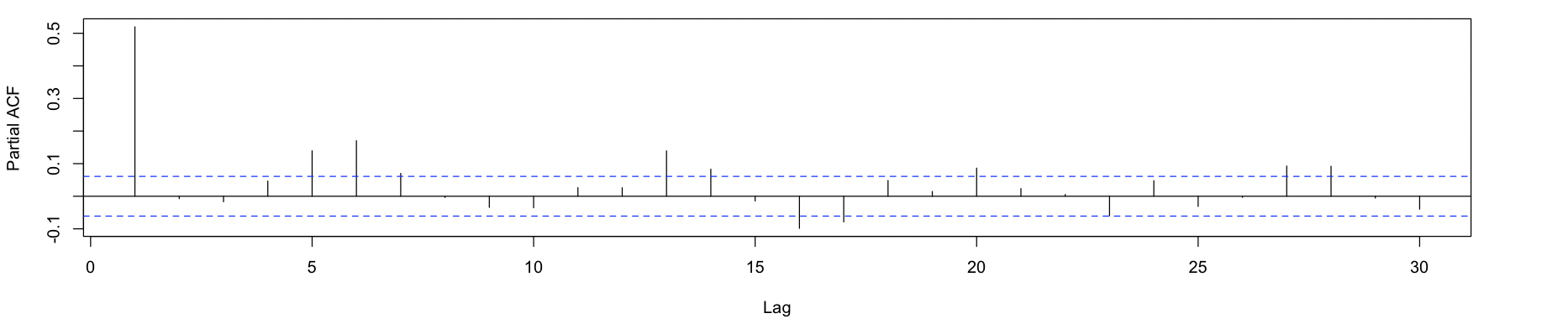}
  \caption{Final Residuals: PACF}
  \label{fig:fpacf}
\end{figure}

The cycle removed residuals are shown in Fig. \ref{fig:fresid} and it looks like they are quite free from any trends or cycles as such and look more like white noise. ADF test on these residuals has a p-value of 0.001 which shows the left out residuals are much more stationary with higher confidence. It was observed that removal of these 5 cycles introduced some other period cycles as shown in Fig. \ref{fig:cycremov} but those are not removed given their less relative densities. ACF and PACF of these final residuals as shown in Fig. \ref{fig:facf} and Fig. \ref{fig:fpacf} respectively, which strengthen the hypothesis that AR(1) might be a good fit given ACF is tailing off much better as compared to raw data and PACF still cuts off at lag 2.

\section{Modelling Approaches}
From the ACF and PACF plots it looks like the final residuals can be modelled by an AR(1) time series model. Also ARMA models can be helpful to fit this kind of noisy and complex data given the original time series had both ACF and PACF have somewhat oscillating values. ARIMA models are not considered given the time series is stationary from the beginning and that d parameter would be 0. 

\begin{figure}[h!]
    \centering
    \subfigure[]{\includegraphics[scale=0.40]{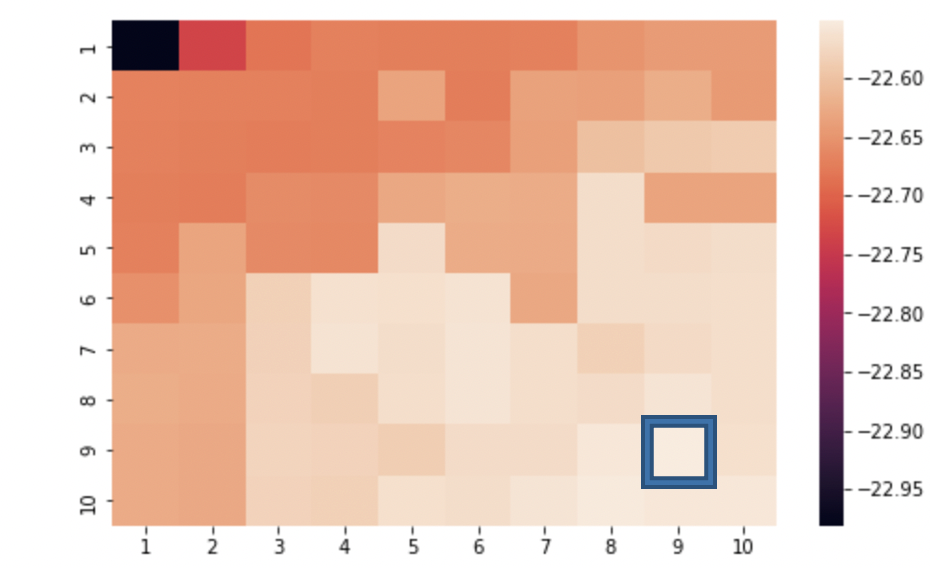}} 
    \hspace{0.5cm}
    \subfigure[]{\includegraphics[scale=0.40]{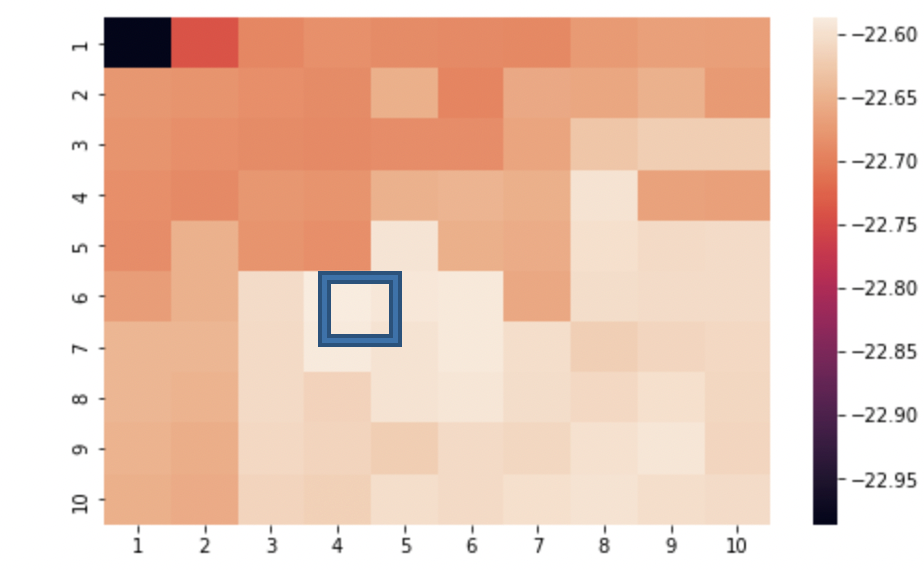}}
    \caption{(a) AIC scores  (b) Harmonic mean of AIC and BIC scores}
    \label{fig:abscores}
\end{figure}

To find the potential p and q parameters for the ARMA(p, q) model, grid search is used with p, q $\in [1, 10]$ where p is the order of Autoregressive (AR) model and q is the order of Moving Average (MA) model. p and q values are restricted below 10 to avoid overfitting and that runtimes are very high for higher order models. Akaike Information Criterion (AIC) and Bayesian Information Criterion (BIC) scores are calculated for every model. The AIC scores are shown as a heatmap in Fig. \ref{fig:abscores}(a) which supports ARMA(9, 9) as a good candidate with lowest AIC score of 22.55. From the heatmap it is observed that variance of AIC scores is very low and choosing such a high order model can lead to overfitting. To avoid this, harmonic mean of AIC and BIC scores is plotted, as shown in Fig. \ref{fig:abscores}(b) and ARMA(6, 4) is chosen as another potential candidate based on lowest harmonic mean score of 22.58. BIC score penalizes model complexity and thus helps in choosing simpler models.

\section{Results and Discussion}
\begin{figure}[h!]
  \centering
  \includegraphics[scale=0.3]{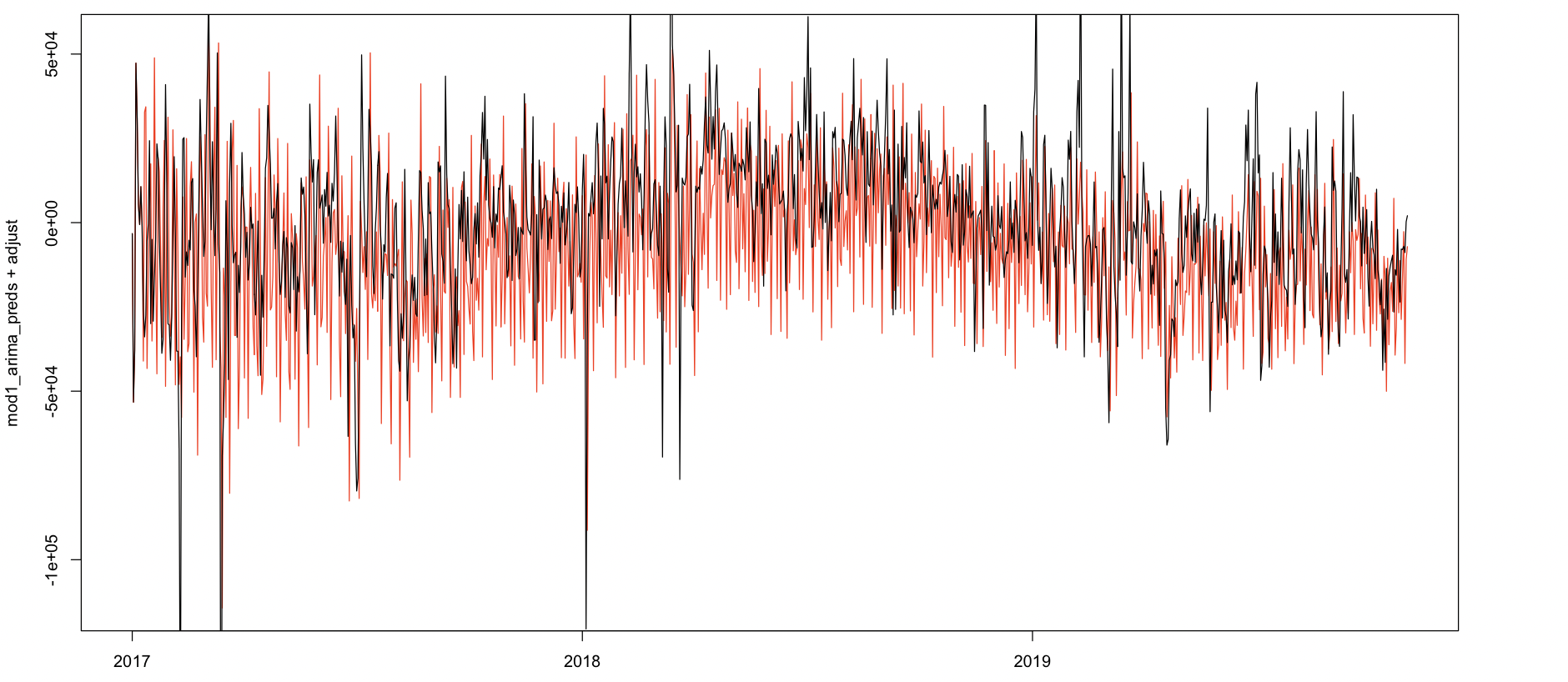}
  \caption{ARMA(9, 9): Train Residuals prediction (red) without level adjustment }
  \label{fig:mod11}
\end{figure}

\begin{figure}[h!]
  \centering
  \includegraphics[scale=0.3]{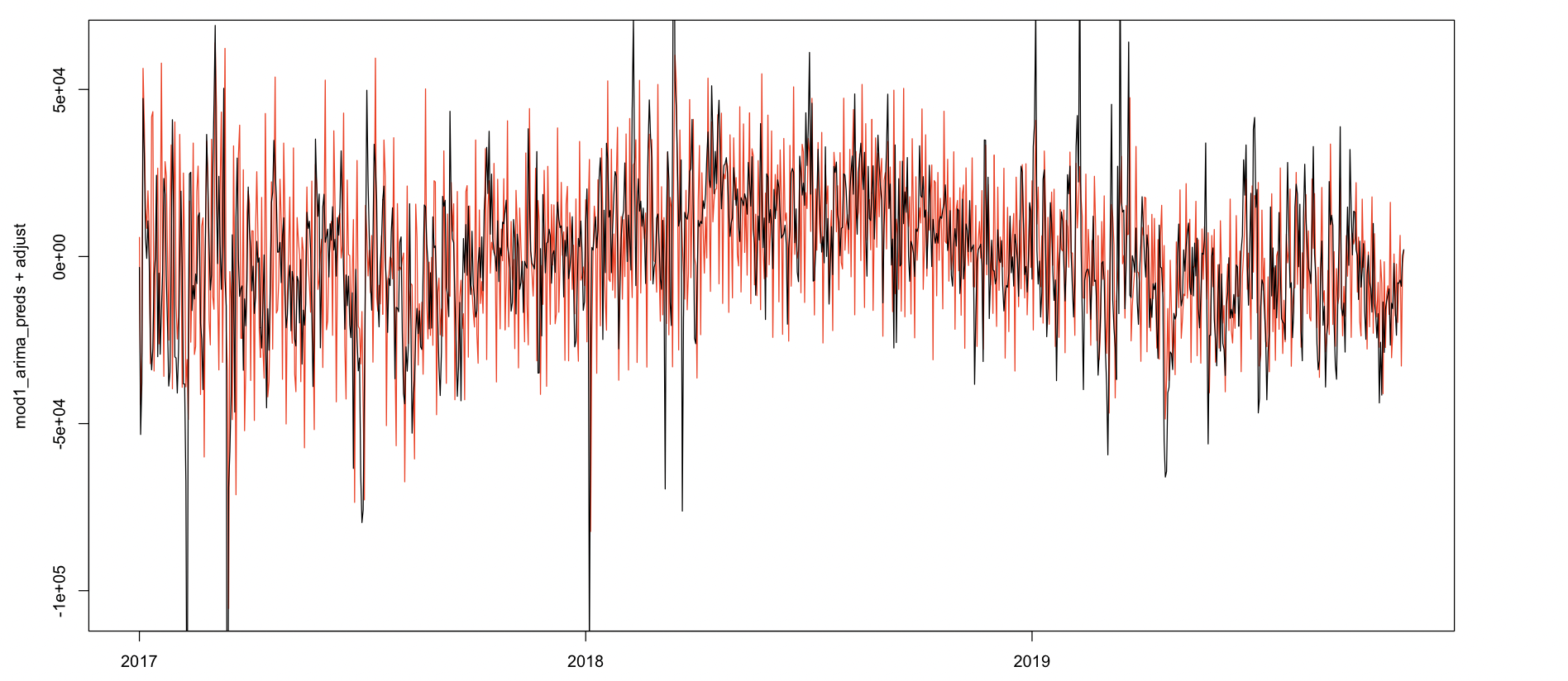}
  \caption{ARMA(9, 9): Train Residuals prediction (red) with level adjustment}
  \label{fig:mod12}
\end{figure}

\begin{figure}[h!]
  \centering
  \includegraphics[scale=0.3]{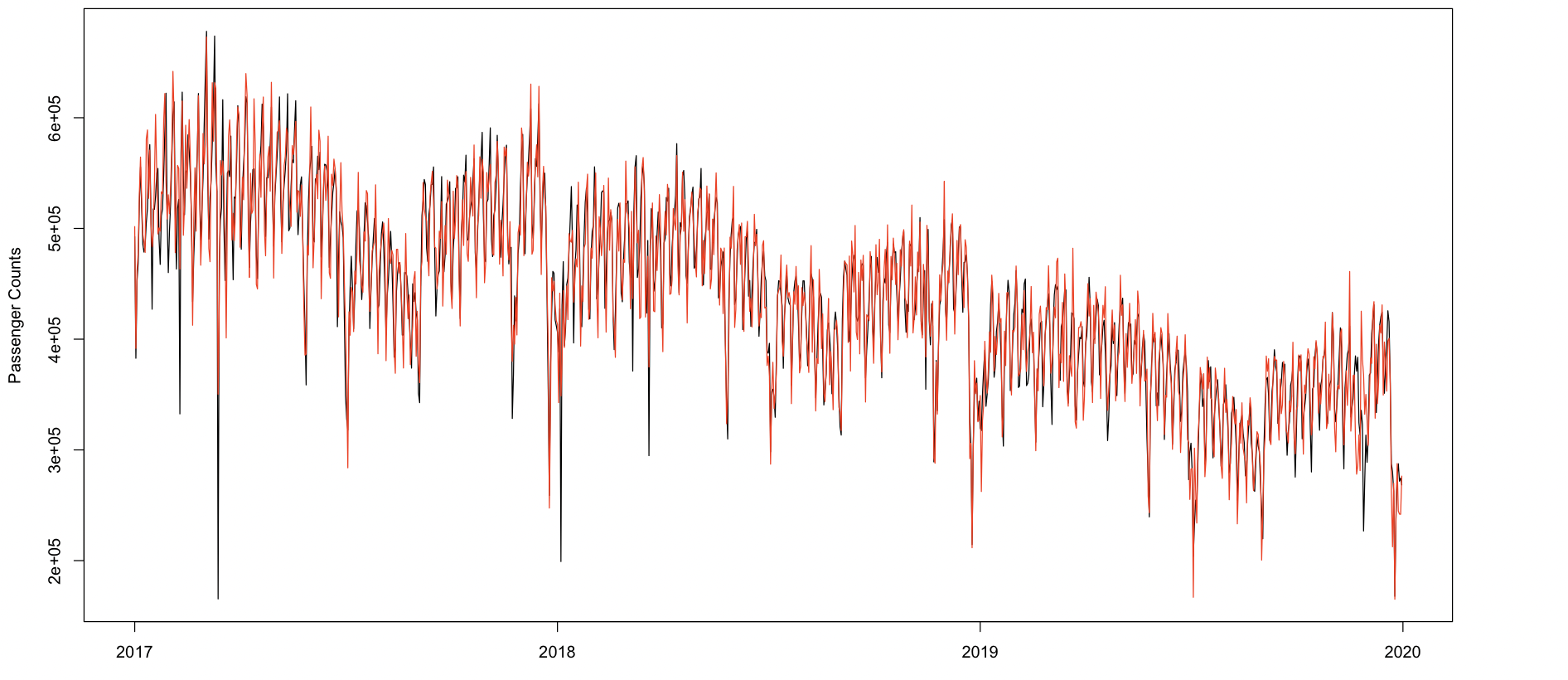}
  \caption{ARMA(9, 9): Train data predictions (red) with level adjustment}
  \label{fig:mod13}
\end{figure}

\begin{figure}[h!]
  \centering
  \includegraphics[scale=0.3]{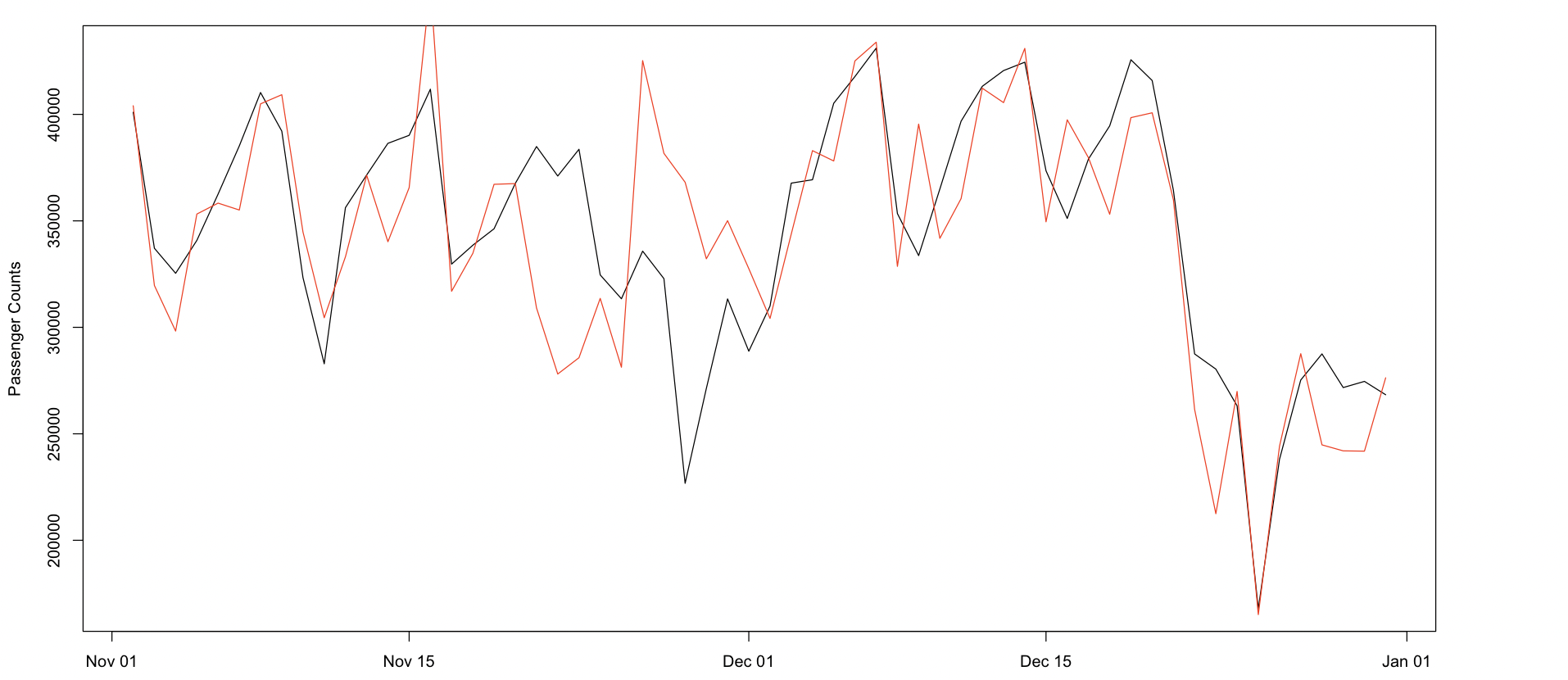}
  \caption{ARMA(9, 9): Test data predictions (red) with level adjustment}
  \label{fig:mod14}
\end{figure}

The first model, ARMA(9, 9), is fitted on final train residuals and is shown in Fig. \ref{fig:mod11} which is extremely choppy with sudden spikes. The train residual RMSE is 27848.66. One thing noticed is that, the shape of the predictions is matching with the actual residuals but there is some positive constant missing which makes the prediction values small. One of the possible reasons for this anomaly is that it's a complex model and relies a lot on the first 9 values. Observing Fig. \ref{fig:mod11}, it looks like the model did poorly in the start which gave the predictions a poor momentum, making the further predictions underwhelming by some constant. To mitigate this, different positive constants are added to the predictions till the train RMSE decreased. A constant of +9000 is added and the residual predictions are shown in Fig. \ref{fig:mod12} which decreased train RMSE by 1390.2. The full train data predictions are shown in Fig. \ref{fig:mod13} which again is quite discontinuous with sharp peaks and looks like an overfit with RMSE of 26458.46. The test predictions on last two months of hold data is shown in Fig. \ref{fig:mod14} and has a RMSE of 39954.88.

\begin{figure}[h!]
  \centering
  \includegraphics[scale=0.3]{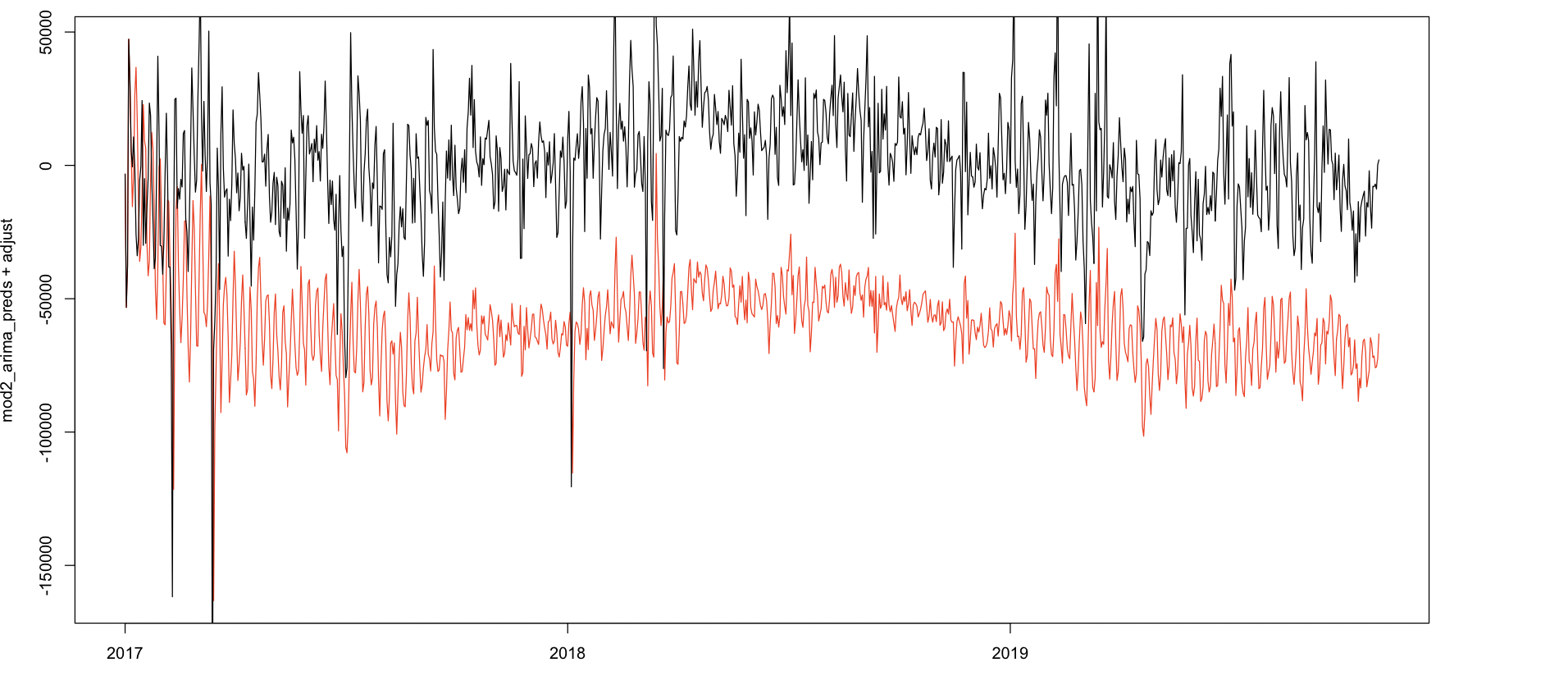}
  \caption{ARMA(6, 4): Train Residuals prediction (red) without level adjustment }
  \label{fig:mod21}
\end{figure}

\begin{figure}[h!]
  \centering
  \includegraphics[scale=0.3]{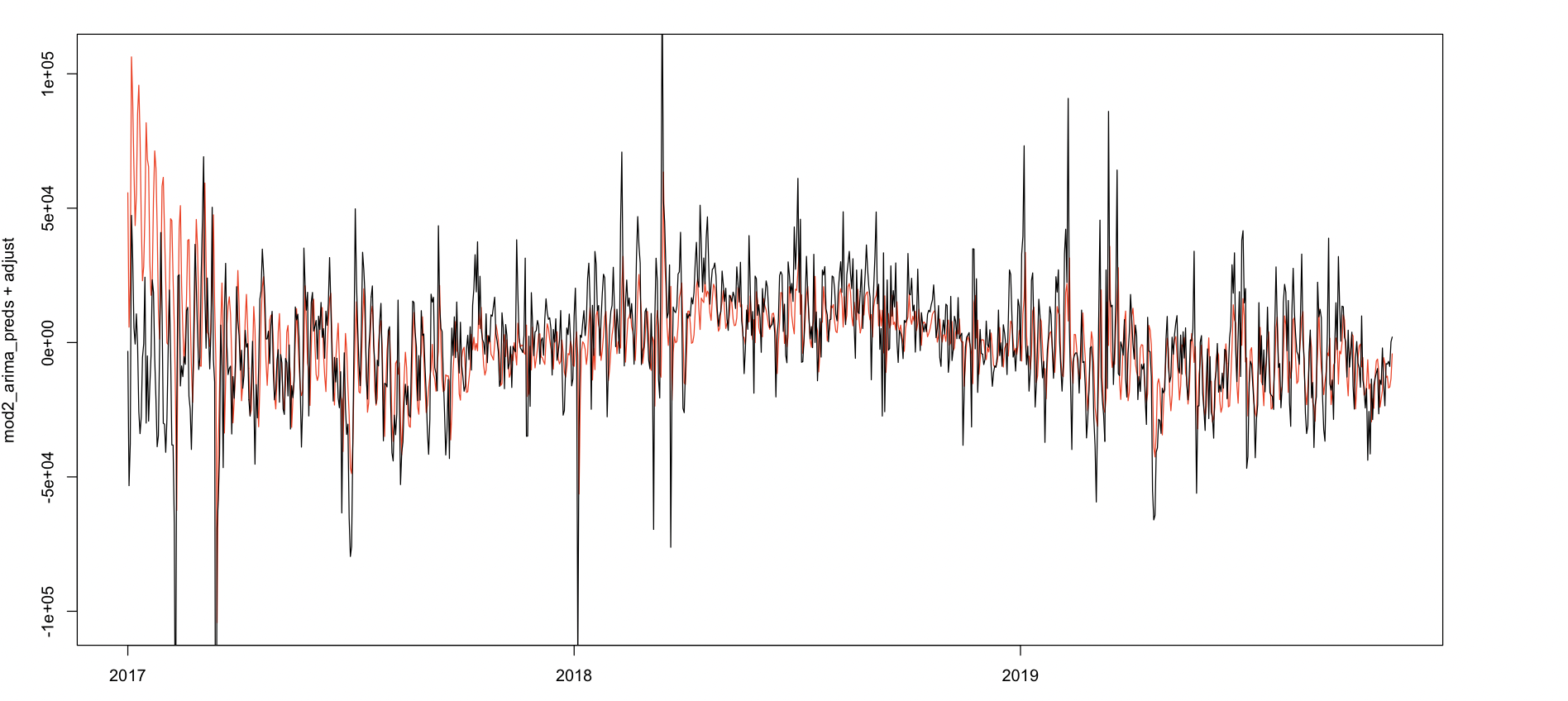}
  \caption{ARMA(6, 4): Train Residuals prediction (red) with level adjustment}
  \label{fig:mod22}
\end{figure}

\begin{figure}[h!]
  \centering
  \includegraphics[scale=0.3]{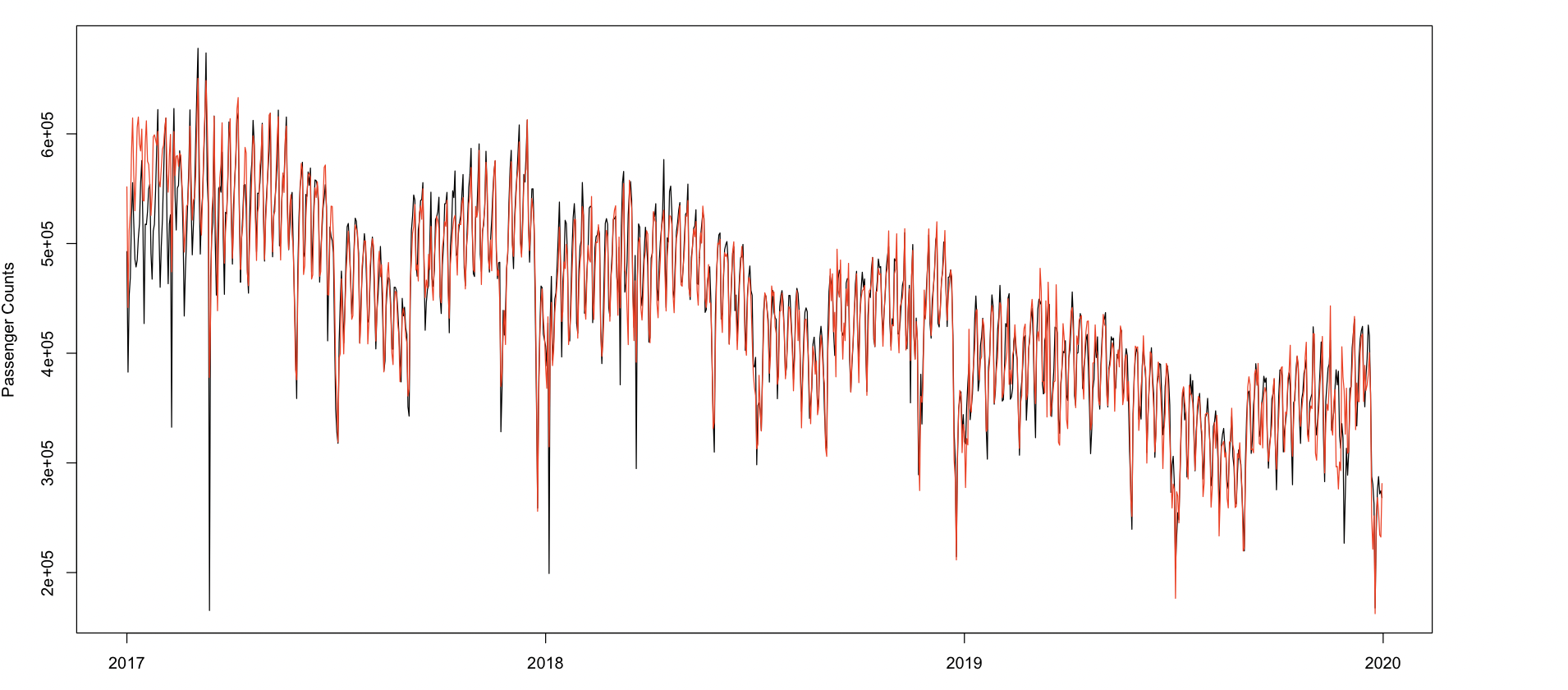}
  \caption{ARMA(6, 4): Train data predictions (red) with level adjustment}
  \label{fig:mod23}
\end{figure}

\begin{figure}[h!]
  \centering
  \includegraphics[scale=0.3]{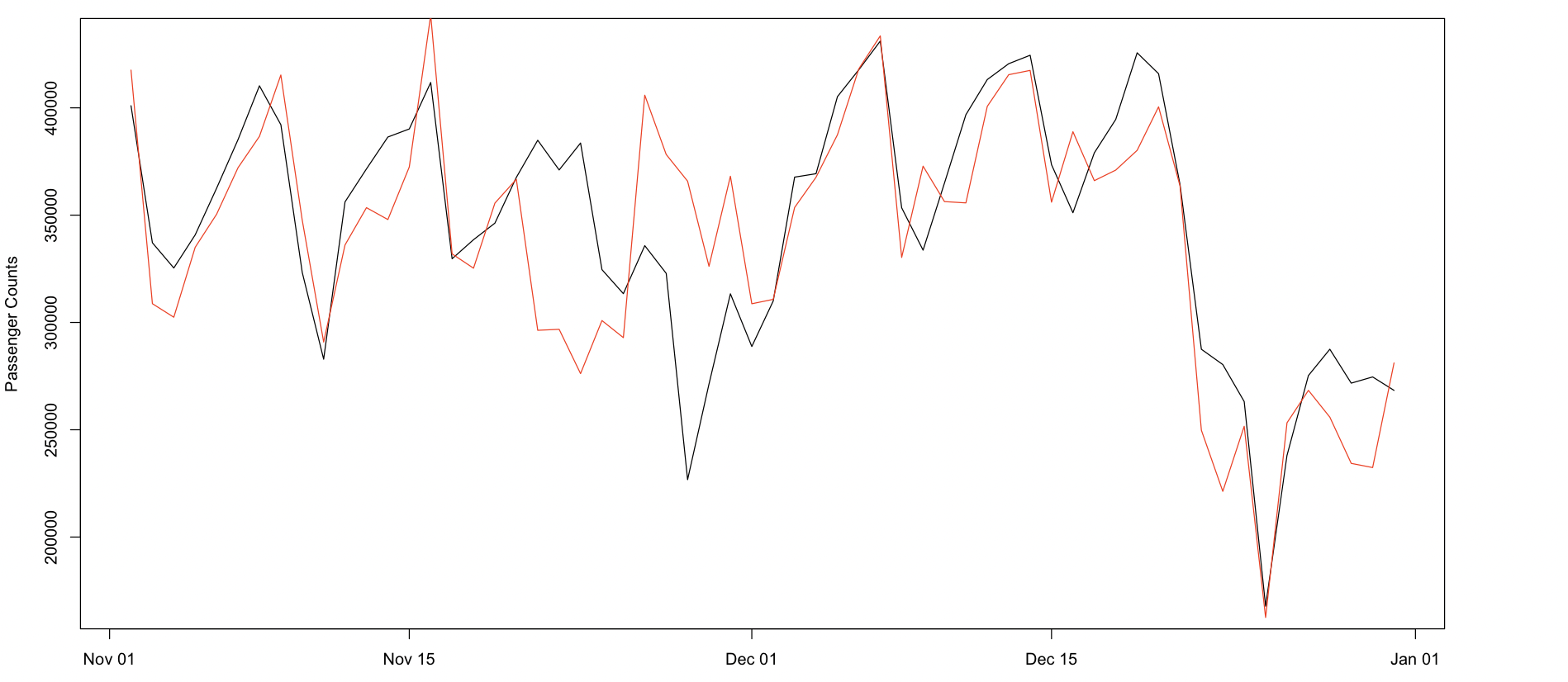}
  \caption{ARMA(6, 4): Test data predictions (red) with level adjustment}
  \label{fig:mod24}
\end{figure}

The second model, ARMA(6, 4) is simpler compared to first one and the fit on train residuals is shown in Fig. \ref{fig:mod21} which looks a bit smooth and is less over-fitting when compared to ARMA(9, 9). The train residual RMSE is 63430.52 which is quite high but looking at predictions in Fig. \ref{fig:mod21}, the shape of the prediction is quite good, it's just the issue of a missing constant. A positive constant of +59000 is added to reduce the RMSE by 40348.29 and the final residual fit is shown in Fig. \ref{fig:mod22}. The reasoning done for ARMA(9, 9) about why this happens still holds here, given this is also a complex model but the constant required here is quite large. One of the reasons for this is that the AR coefficients(0.02, 0.08, 0.04...) of ARMA (9, 9) are small thus they offered less shift and wrong momentum as compared to ARMA(6, 4) which has large AR coefficients (2.31, -2.32, 0.97...) thus drifting the start by a large error. The full train data predictions are shown in Fig. \ref{fig:mod23} which again is less discontinuous as compared to ARMA(9, 9) but still looks like an overfit with RMSE of 23082.23. The test predictions on last two months of hold data is shown in Fig. \ref{fig:mod24} has a RMSE of 37890.48 which is an improvement over ARMA(9, 9) showing that simpler models for this data are working better.

\begin{figure}[h!]
  \centering
  \includegraphics[scale=0.3]{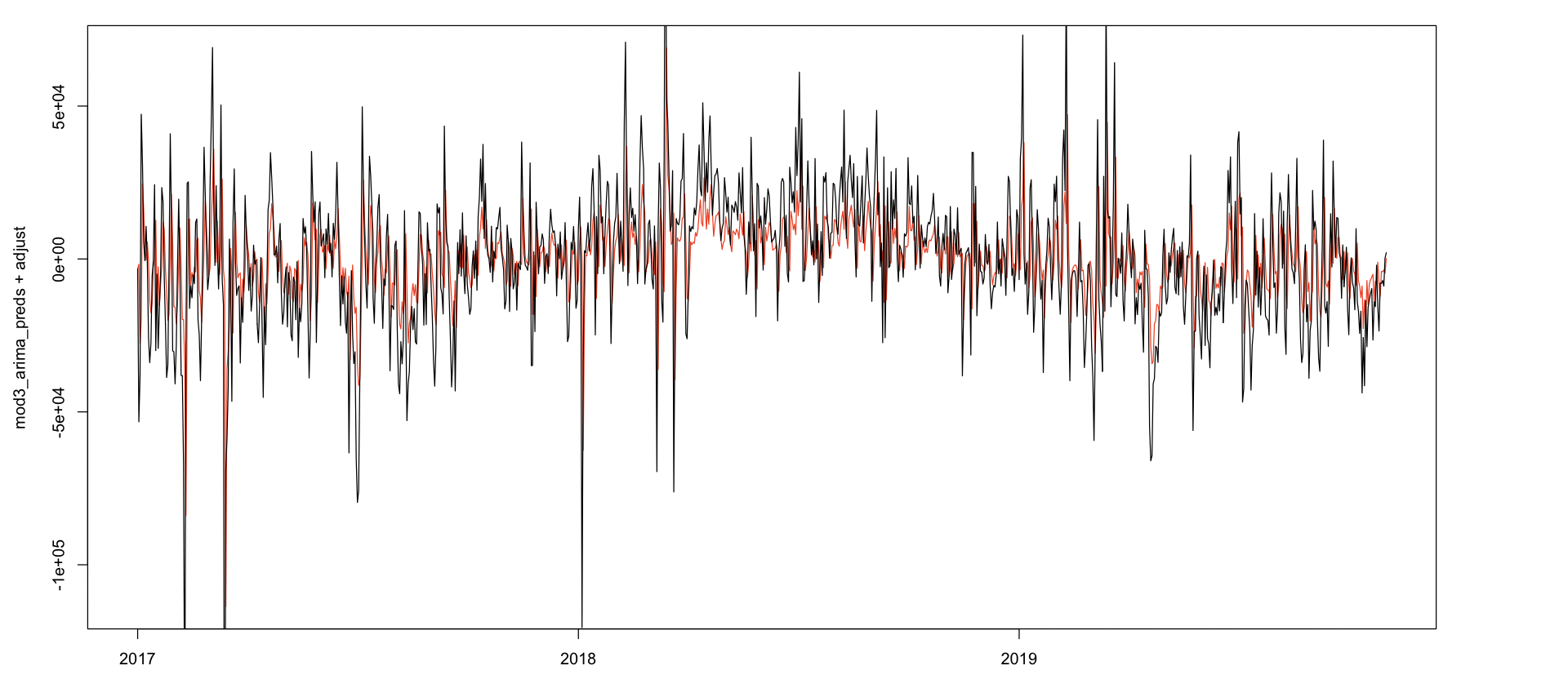}
  \caption{ARMA(1, 0): Train Residuals prediction (red)}
  \label{fig:mod31}
\end{figure}

\begin{figure}[h!]
  \centering
  \includegraphics[scale=0.3]{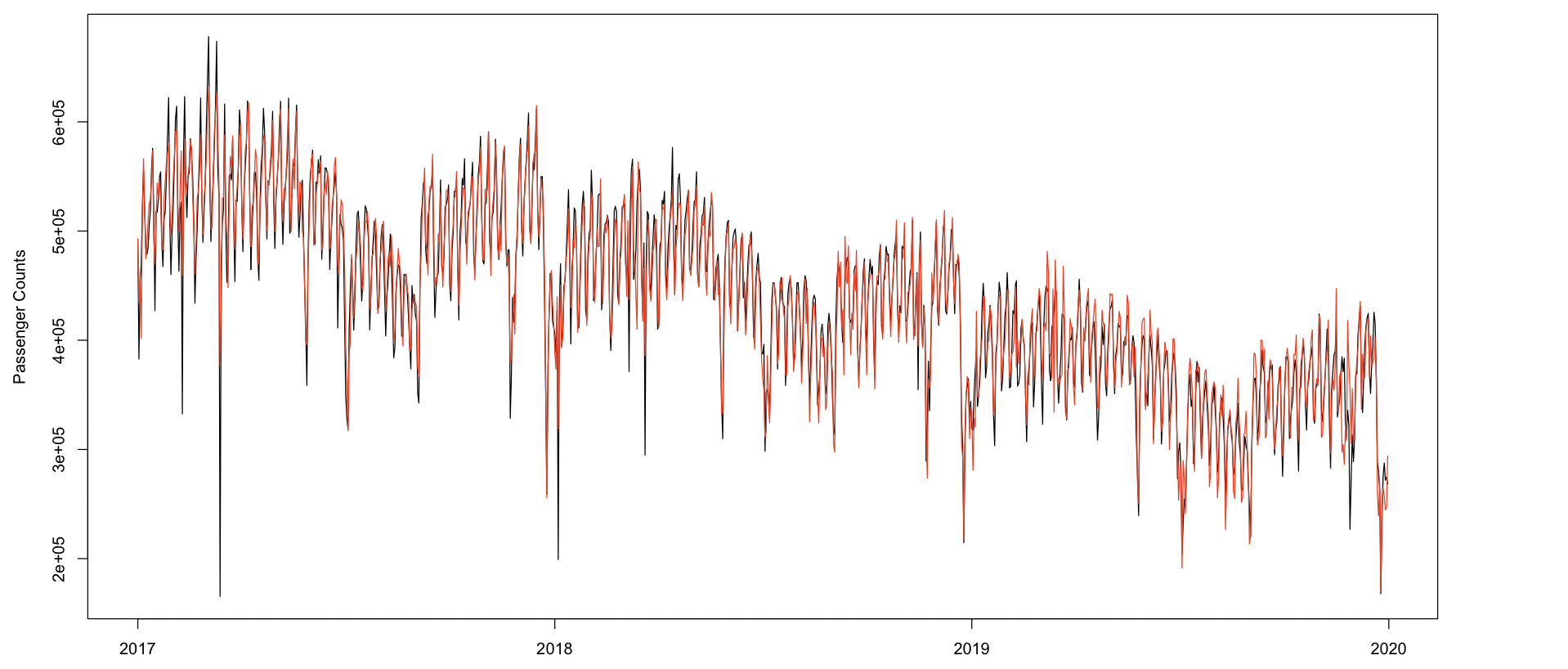}
  \caption{ARMA(1, 0): Train data predictions (red)}
  \label{fig:mod32}
\end{figure}

\begin{figure}[h!]
  \centering
  \includegraphics[scale=0.3]{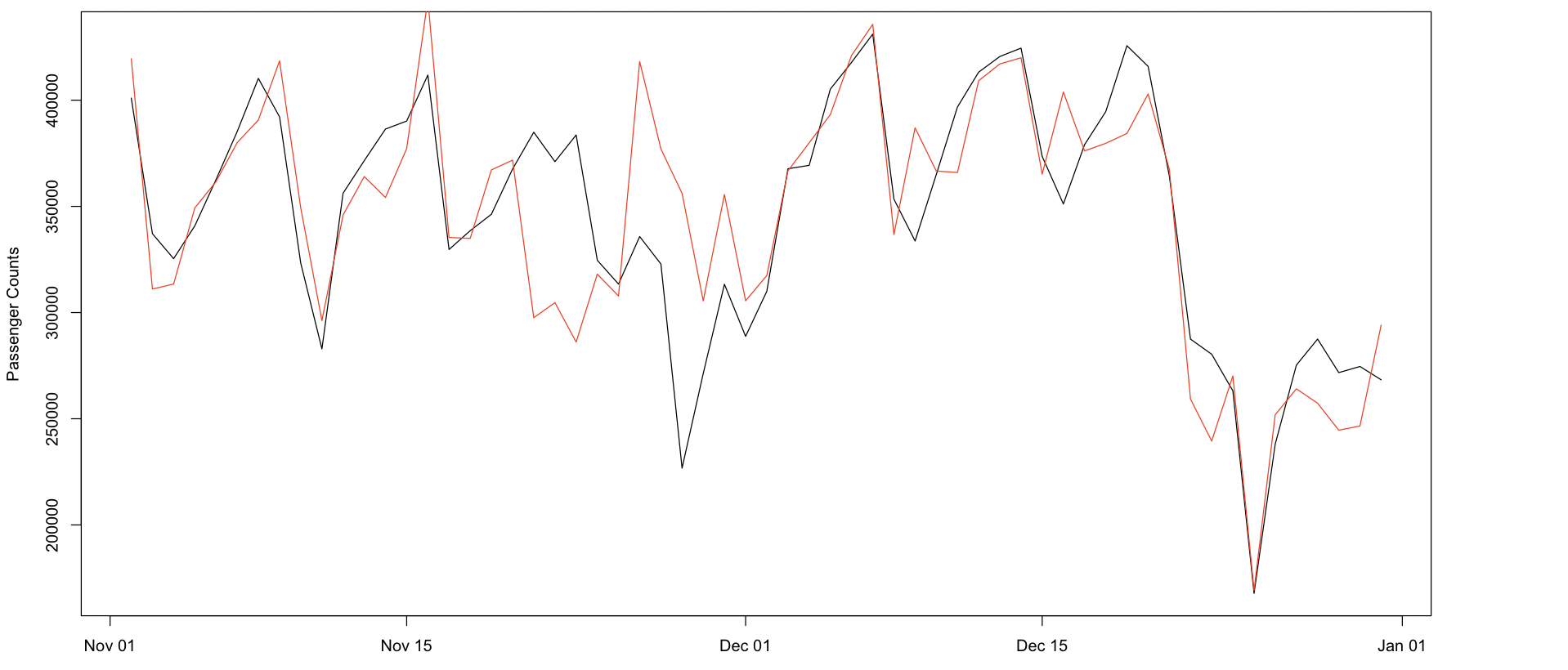}
  \caption{ARMA(1, 0): Test data predictions (red)}
  \label{fig:mod33}
\end{figure}

The simplest model ARMA(1, 0), which is found from the ACF and PACF plots earlier, gets the best results. The residual predictions are smoother as compared to previous complex models with less signs of overfitting as shown in Fig. \ref{fig:mod31} and has a RMSE of 20177.38. The best part is that there is no need for level adjustment given adding or removing a constant from the predictions increases the RMSE. The train data predictions fit much better as compared to other models with less sharp peaks and variations as shown in Fig. \ref{fig:mod32} and has a train RMSE of 20177.38 which is the least obtained error. The test RMSE obtained from this model is the smallest with a value of 34880.39 and the test data predictions are shown in Fig. \ref{fig:mod33}. The AR(1) coefficient has a value of 0.519 which is not a large one, explains why the predictions are not continuous and has less effect of $X_{t - 1}$. All the results are consolidated in Table \ref{tab:results}. Given the mean of data is around 438K passengers/day the test RMSE of AR(1) is quite well within the permissible limits.

The question arises that since simplest of the models did well, will there be any further improvements with even more simpler models? The answer to that is NO, for this data. The Table \ref{tab:simpleresults} emphasises on this. Various simpler models like predicting just the train mean for test data, simple linear regression, linear regression with 7-day cycle (having most power) removed, model which gives previous day value as prediction, are tried but the Test RMSEs observed are way too large when compared with AR(1). Also the modelling process involved removing of cycles but is it really needed for this data? The answer to this is YES, its important given it makes the residuals much more stationary making them more suitable for time series modelling. AR(1) model on residuals without any cycle removal has the Test RMSE of 36742.08 which is quite large when compared the best RMSE achieved.

\section{Future Improvements}
\vspace{-0.5cm}
From the test predictions, the AR(1) model did well, achieving quite reasonable RMSE but the results are not that good if the predictions in Fig. \ref{fig:mod33} are analysed in depth. The model doesn't do well if there is a significant difference between the previous days passenger counts.  This can be seen especially in the period of mid-November to the first week of December. ARMA model can't account for the sudden dips and rises which are not seen in history. One approach is that outliers can be removed in the data processing phase and data can be smoothened a bit before modelling. Another thing is that daily passenger counts depends a lot on different factors like the day, date and weekday of the year, if there is some change in local laws and regulations, daily temperatures, domestic/foreign travel policies given NYC is a tourist place, etc. Thus a better approach for this time series would be to use Multi-variate time series modelling to account for better logical variations in the data and in turn better predictions.


\printbibliography

\vspace{6cm}
\section*{Appendix}

\begin{table}[h!]
\centering
 \begin{tabular}{||c| c| c| c| c||} 
 \hline\hline
 & Estimate & Std. Error & t value & Pr(> |t|) \\ [0.5ex]
 \hline
 Intercept & 547356.006 & 3401.829 & 160.90 & <2e-16 *** \\ 
 Time & -200.138 & 5.689 & -35.18 & <2e-16 *** \\
 \hline
 \end{tabular} \\ [1ex] 
 \caption{Linear Trend Summary}
 \label{tab:lm}
\end{table}

\begin{table}[h!]
\centering
 \begin{tabular}{||c| c| c| c||} 
 \hline\hline
 Model(p, d, q) & Adjustment & Train RMSE & Test RMSE \\ [0.5ex]
 \hline
 ARIMA(9, 0, 9) & +9000 & 26458.46 & 39954.88 \\ 
 ARIMA(6, 0, 4) & +59000 & 23082.23 & 37890.48\\
 ARIMA(1, 0, 0) & 0 & 20177.38 & 34880.39\\
 \hline
 \end{tabular} \\ [1ex]
 \caption{Model Evaluations}
 \label{tab:results}
\end{table}

\vspace{-1cm}
\begin{table}[h!]
\centering
 \begin{tabular}{||c| c||} 
 \hline\hline
 Model & Test RMSE \\ [0.5ex]
 \hline
 Mean of Train Data & 113420.61\\ 
 Linear Regression(LR) & 56739.59\\
 LR + 7-day cycle & 52615.37\\
 $X_{t} = X_{t - 1}$ & 38643.50\\
 AR(1) without cycle removal & 36742.08\\
 \hline
 \end{tabular} \\ [1ex]
 \caption{Comparison with simpler models}
 \label{tab:simpleresults}
\end{table}

\end{document}